\documentclass[10pt, conference,letterpaper]{IEEEtran}
\usepackage{graphicx}
\usepackage{algorithmic}
\usepackage{algorithm}
\usepackage{amsmath}
\usepackage{booktabs}
\usepackage{multirow}
\usepackage{array}
\usepackage{setspace}

\ifCLASSINFOpdf
\else
\fi
\begin{document}
%
\title{An Intelligent Indoor Positioning Algorithm Based on Wi-Fi and Bluetooth Low Energy}
%
%
%

\author{Karamat~Beigi\IEEEmembership{Student Member,~IEEE}
        and~Hamed~Shah-Mansouri~\IEEEmembership{Member,~IEEE}
		  \\ 
    Department of Electrical Engineering, Sharif University of Technology, Tehran, Iran \\
    email:  keramat.beygi@ee.sharif.edu, hamedsh@sharif.edu
    
}

\maketitle

\begin{abstract}
Indoor positioning plays a pivotal role in a wide range of applications, from smart homes to industrial automation. In this paper, we propose a comprehensive approach for accurate positioning in indoor environments through the integration of existing Wi-Fi and Bluetooth Low Energy (BLE) devices. The proposed algorithm involves acquiring the received signal strength indicator (RSSI) data from these devices and capturing the complex interactions between RSSI and positions. To enhance the accuracy of the collected data, we first use a Kalman filter for denoising RSSI values, then categorize them into distinct classes  using the K-nearest neighbor (KNN) algorithm. Incorporating the filtered RSSI data and the class information obtained from KNN, we then introduce a recurrent neural network (RNN) architecture to estimate the positions with a high precision. We further evaluate the accuracy of our proposed algorithm through testbed experiments using ESP32 system on chip with integrated Wi-Fi and BLE. The results show that we can accurately estimate the positions with an average error of 61.29 cm, which demonstrates a 56\% enhancement compared to the state-of-the-art existing works.
\end{abstract}


%
\IEEEpeerreviewmaketitle

\section{Introduction}
%
%
%
%
\IEEEPARstart{I}{ndoor} positioning is the process of determining the location of a person or object inside a building or other indoor spaces. Indoor positioning technologies have various applications across various industries, such as retail \cite{Access23}, healthcare \cite{Access27}, warehousing \cite{Access24}, manufacturing \cite{Access25}, airports \cite{Access26}, and more. Although the global positioning system (GPS) has been widely used to determine the location of an object or person in outdoor environments, signals are often blocked or weakened in indoors, making it difficult to be used for positioning \cite{Access22}. Several technologies can be used for indoor positioning such as Wi-Fi, Bluetooth Low Energy (BLE) and Zigbee among which Wi-Fi and BLE are widely used due to their ubiquitous availability in most indoor environments, compatibility with devices, low power consumption, and cost effectiveness. In addition, Wi-Fi and BLE-based positioning systems can be easily scaled to cover large indoor areas, which make them suitable for use in large buildings such as airports and shopping malls.

\begin{figure}[!t] 
\centering
\includegraphics[width=.3\textwidth]{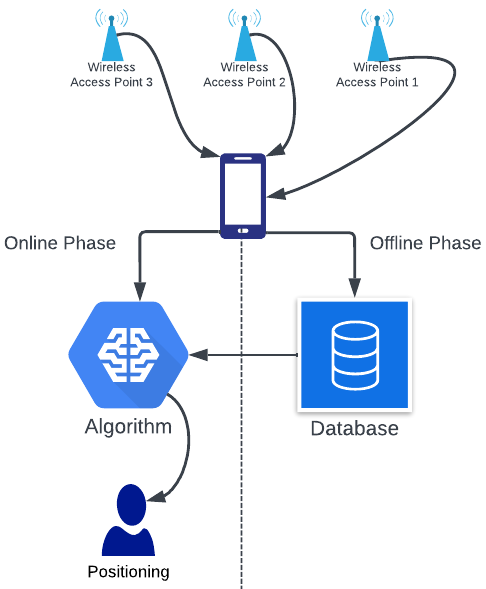}
\DeclareGraphicsExtensions.
\caption{Fingerprinting method.}
\vspace{-5mm}
\label{fig:fingerprinting}
\end{figure}

Several indoor positioning techniques are available such as trilateration, proximity-based, and fingerprinting. In this work, we focus on fingerprinting technique \cite{Access20} as it does not need extra devices to be deployed in the environment and can still achieve high accuracy. 
Fingerprinting refers to a promising indoor positioning technique that entails meticulously mapping unique signal patterns, including Wi-Fi and Bluetooth, across specific locations. These distinctive patterns, referred to as fingerprints, serve as fundamental references for precisely determining a device's position within indoor spaces. The process involves online and offline phases, as shown in Fig. \ref{fig:fingerprinting}.
In the offline phase, the system collects data from various access points (APs) already present in the room. This data includes received signal strength indicators (RSSI), unique identifiers, and coordinates that can be used to identify the location of an object or person. This data is collected at different locations in the room and stored in a database for later use. In the online phase, the system uses the data collected during the offline phase to determine the location in real-time. When an object or person enters a room or moves around inside a building, the system compares its RSSI with those stored in its database to determine the location.
The accuracy of the fingerprinting indoor positioning depends on several factors, such as the number of APs, their placement, and environmental factors, such as interference from other wireless devices. However, with proper planning and intelligent decision making, fingerprinting indoor positioning can effectively locate people and objects inside buildings.
\subsection{Related Work}
     Indoor positioning is an established field with many innovations and techniques being explored. An overview of the subject can be found in \cite{Access6}, \cite{Access21}. The authors in   \cite{Access15} provided a comprehensive analysis of an indoor positioning system using Wi-Fi and fingerprinting, along with machine learning techniques. A technique called high adaptivity indoor positioning (HAIL) is presented in \cite{Access10}. The proposed technique utilizes absolute and relative RSSI values, along with a backpropagated neural network. A simulation-based location tracking system using Wi-Fi and fingerprinting is proposed in \cite{Access11}. The authors developed a fuzzy matching algorithm by using a particle swarm optimization. In  \cite{Access12}, the authors analyzed the effect of the beacon's location while considering the effect of the grid size and the location of the reference points (RPs). Using a K-means algorithm, the authors in \cite{Access13} proposed an efficient method for fingerprint-based positioning of multi-story buildings. Although RSSI measurement has been widely used for fingerprinting-based indoor positioning, RSSI values are volatile as a result of non-line-of-sight (NLOS) and multipath propagation in the wireless channel, as well as hardware heterogeneity. The authors in \cite{Access14} proposed a normalized rank-based support vector machine (SVM) that achieves room-level accuracy and aims to mitigates RSSI fluctuations.
\subsection{Motivation and Contributions}
The aforementioned studies have proposed various indoor positioning techniques. However, an efficient technique with high accuracy is yet to be explored. Unlike outdoor environments, indoor spaces have more obstacles, walls, and materials that can block or reflect signals from Wi-Fi, Bluetooth, or other positioning technologies. This can lead to signal attenuation and interference, making the design an accurate algorithm more challenging.

In this paper, we propose a fingerprinting-based indoor positioning algorithm. The experimentation occurs within a room environment where we optimally place the Wi-Fi AP and BLE beacons in the room. Our proposed algorithm acquires the RSSI values and exploits the advantages of machine learning to accurately estimate the position of the device. In particular, we use K-nearest neighbors (KNN) to categorize the RSSI values into distinct classes and label them. These will be then fed to a recurrent neural network (RNN) model. 
The main contributions of this paper are summarized as follows.
\begin{itemize}
\item \textit{Problem Formulation}: We formulate an optimization problem to determine the optimal placement of WiFi APs and BLE beacons. The problem aims to maximize the overall dissimilarity in RSSI values among the RPs.
    \item \textit{Data Acquisition and Preprocessing }: We acquire RSSI values using ESP32 devices. To overcome the RSSI noisy measurements and fluctuations, we design a Kalman filter to smooth out the RSSI values. We then classify these acquired values into distinct classes and label them using the KNN algorithm. This data preprocessing further enhances the accuracy of the position estimation. 
    \item \textit{RNN-based Algorithm Integration}: We then design an RNN model. Once trained, the RNN-based algorithm can make accurate position estimations for unseen data. Given a sequence of RSSI measurements from the APs and  BLE beacons, the RNN estimates the device position by exploiting the learned patterns, thereby offering enhanced accuracy and robustness. 
    \item \textit{Testbed Experiments}: We develop a testbed using ESP32 system on chip, which incorporates Wi-Fi and BLE, to collect real-world traces for the purpose of training and validating our algorithm. We then evaluate the accuracy of our proposed algorithm. We conduct thorough experiments using real-world data and compare the estimated positions with ground truth values. The results demonstrate a notable enhancement in location estimation obtained by our proposed algorithm with an average error of 61.29 cm. This exhibits a 56\% improvement compared to the existing work \cite{Access17}.
\end{itemize}
This paper is organized as follows. In Section II, we present the experimental procedure and data collection. In Section III, we present the position estimation using neural
networks. In Section IV, we evaluate the performance of the algorithm. Finally, Section V concludes the paper.

\section{Experimental Procedure and Data Collection}
In this section, we present the meticulous experimental methodology undertaken to acquire the requisite data for the design and validation of our indoor positioning algorithm. The experimental environment comprises a room with dimensions of 7 meters in length and 4 meters in width. Within this environment, two BLE beacons and one Wi-Fi AP are available and used as the signal sources for fingerprinting-based positioning, as depicted in Fig. 2. 
\subsection{Beacons and AP placement}
\begin{figure}[!t]
\centering
\includegraphics[width=.35\textwidth]{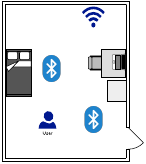}
\DeclareGraphicsExtensions.
\caption{Experimental environment.}
\label{exp_enviroment}
\vspace{-5mm}
\end{figure}
Fingerprint-based positioning suggests that when the radio map contains highly similar fingerprints, accurately identifying the best-matched fingerprint for a given RSSI becomes challenging. To enhance positioning accuracy, it is imperative to minimize the presence of similar fingerprints in the radio map. The goal is to position the beacons and AP strategically to maximize the dissimilarity between fingerprints of neighboring RPs. The similarity between fingerprints is determined by their Euclidean distance, and increasing this distance effectively amplifies dissimilarity, leading to improved positioning. These fingerprints are derived from a selected subset of beacons and APs among all available RPs. This approach involves placing signal sources in appropriate locations, eliminating reliance on actual RSSI measurements. Predicting the device location from RSSI values is challenging due to indoor radio propagation complexities influenced by multipath fading and shadowing caused by obstacles. We employ the log-distance path loss model  (\ref{eq:rrsi}) to determine the RSSI values at various distances from the signal source. $RSSI_0$ is the reference signal strength at a known distance $d_0$, and  $\alpha$ is the path-loss exponent, signifying how rapidly the signal strength diminishes with distance. The constant $d_0$ remains fixed, while $\alpha$ is determined through empirical data in practical applications.
\begin{equation} \label{eq:rrsi}
RSSI = RSSI_0 - 10\alpha\log_{10}\left(\frac{d}{d_0}\right)
\end{equation}
In order to place the AP and beacons, we aim to maximize the overall dissimilarity in RSSI values among the RPs. The number of RPs, denoted as $p$, is determined based on the granularity of these points and the size of the area of interest. The dissimilarity between $RP_i$ and $RP_j$ is measured by the Euclidean distances calculated from the RSSI values and is denoted as:
\begin{equation*}
E(\Theta_i, \Theta_j) = \left( \sum_{l=1}^{k} (RSSI_{l,i} - RSSI_{l,j})^2\right)^{1/2}  
\end{equation*}
where $k$ corresponds to the number of signal sources involved in the positioning system. We now formulate the optimization problem as follows:
\begin{subequations} 
\label{eq:OurProblem}
\begin{flalign}
\text{maximize} \quad &\sum_{i=1}^{p}\sum_{j\in \mathcal{A}_t} E(\Theta_i, \Theta_j) \label{eq:objective} \\
\text{subject to} \quad & \mathcal{A}_t = \{ j \,|\, ||RP_i,RP_j||_2^{1/2} < r \} \label{eq:constraintA} \\
& RSSI_{l,i} = RSSI_0 - 10\alpha \log_{10}\left(\frac{d}{d_0}\right) \label{eq:constraintRSSI} \\
& d = ||(x_i, y_i) - (x_l,y_l) ||_2^{1/2} \label{eq:constraintD}\\
& (x_l, y_l) \in D,  l = 1, 2, 3. \label{eq:constraintXY}
\end{flalign}
\end{subequations}
where set $\mathcal{A}_t$, defined in Constraint (\ref{eq:constraintA}), comprises those RPs whose geometric distances from a particular RP (i.e., $RP_i$) are smaller than a constant $r$. This helps define a neighboring area around each RP, capturing nearby points. In (\ref{eq:constraintRSSI}) and (\ref{eq:constraintD}), $RSSI_{l,i}$ represents the RSSI value of signal source $l$ at $RP_i$, which is predicted using the log-distance path loss model. The geometric distance between $RP_i$ and $l^\text{th}$ signal source is computed based on their coordinates, ($x_i$,$y_i$) and ($x_l$,$y_l$). Finally, constraint (\ref{eq:constraintXY}) relates to the physical layout of the signal sources within the area of interest, denoted as $D$. It ensures that the location coordinates  ($x_l$,$y_l$) of each signal source are constrained within this specified area.
We employ particle swarm optimization (PSO) as a powerful tool to address the critical challenge of optimizing the placement of signal resources.
\subsection{Data Acquisition Setup}
ESP32 is a powerful and versatile system on chip that comes with built-in Wi-Fi and BLE capabilities. This makes it an ideal choice for beacons that can transmit signals over both technologies, allowing for more accurate and reliable positioning. The data collection process commenced by deploying a device equipped with Bluetooth and Wi-Fi. To collect the RSSI values, we prepare a Python script that scans for the signals transmitted by the beacons and records the corresponding RSSI values. This can be achieved using libraries like BluePy \cite{Access28} for BLE beacons, or Scapy \cite{Access30} for Wi-Fi APs. We move this device to predetermined locations within the room, spanning its entirety, while recording RSSI values from the BLE beacons and the Wi-Fi AP. The spatial coordinates of each RP are meticulously measured and recorded to facilitate subsequent analysis and evaluation. The process is meticulously repeated to encompass a comprehensive representation of the room's spatial layout. Once the RSSI values are collected, they can be stored in the database, allowing us to preprocess and use data for training the model.
\begin{algorithm}[t]
\caption{RSSI Collection and Kalman Filtering}
\label{alg:combined_short}
\textbf{Initialization:} ESP32 BLE and Wi-Fi\\
\textbf{Input:} Kalman filter parameters \\
\textbf{Output:} Smoothed RSSI.
\begin{algorithmic}[1]
\STATE Initialize BLE scanner
\STATE Initialize Wi-Fi module and scan settings
\STATE $s \leftarrow \text{init\_state}$
\STATE $c \leftarrow \text{init\_cov}$
\STATE $r \leftarrow$ empty list for smoothed RSSI values
\WHILE{True}
    \FOR{devices (BLE or Wi-Fi)}
        \STATE $rssi \leftarrow$ Read RSSI
        \STATE Add $rssi$ to the respective RSSI list (BLE or Wi-Fi)
        \STATE \% Kalman Filtering
        \STATE $m \leftarrow rssi$
        \STATE \% Predict the next state based on the system dynamics
        \STATE $p \leftarrow s$
        \STATE $pc \leftarrow c + pn$
        \STATE \% Calculate Kalman gain
        \STATE $kg \leftarrow \frac{pc}{pc + mn}$
        \STATE \% Update state and covariance
        \STATE $s \leftarrow p + kg \cdot (m - p)$
        \STATE $c \leftarrow (1 - kg) \cdot pc$ 
        \STATE $r$.append($s$)
    \ENDFOR
\ENDWHILE
\RETURN $r$
\end{algorithmic}
\end{algorithm}
\subsection{Data Preprocessing}
When collecting RSSI values from beacons, it is common to encounter noise and fluctuation in the signal that can affect the accuracy of the collected data. One approach for reducing the impact of this noise is to use a filtering algorithm to smooth out the RSSI values and remove unwanted fluctuations. One popular filtering method is the Kalman filter as shown in Algorithm 1, which uses a series of measurements to estimate the state of a system and predict future values. Compared to other filtering methods such as Fourier filtering, the Kalman filter has several advantages in indoor positioning systems. First, the Kalman filter is able to adapt to changes in the environment and adjust its parameters accordingly, while Fourier filtering uses a fixed frequency cutoff that may not be suitable for all situations. This makes the Kalman filter more effective in dynamic indoor environments where the signal characteristics may change over time. Second, the Kalman filter can handle non-linear systems and non-Gaussian noise, which are common in indoor environments. Non-linearities in the signal can arise due to reflections, diffraction, and other environmental factors that affect the signal propagation. Gaussian noise assumptions may not hold in indoor environments due to the presence of interference from other devices, reflections, and other sources of noise. The Kalman filter is better equipped to handle these non-linearities and non-Gaussian noise, providing an accurate estimate of the true RSSI value. Fig. 3 shows the RSSI values before and after applying the Kalman filter. As can be observed, filtering can significantly smooth out the fluctuations.

\section{POSITION ESTIMATION USING NEURAL NETWORK}
As outlined, our work involves employing the RSSI dataset to predict the coordinates between a sender and a receiver through the utilization of artificial neural networks (ANNs). In the first step, we use the KNN to determine which class the online RSSI belongs to, as shown in Fig. 4. In the second step, we use long short-term memory (LSTM) to perform a location estimation task based on RSSI measurements in each predetermined class, as shown in Fig. 5. The employed ANN architecture is designed to address the location estimation task based on RSSI measurements and the training configuration is set as described in Table 1 \cite{Access31}, \cite{Access32} . The ANN utilizes an LSTM layer to capture temporal dependencies in the sequence of RSSI measurements and predict the corresponding position. The architecture consists of the following components:
\begin{figure}[t]
\centering
\includegraphics[width=.45\textwidth]{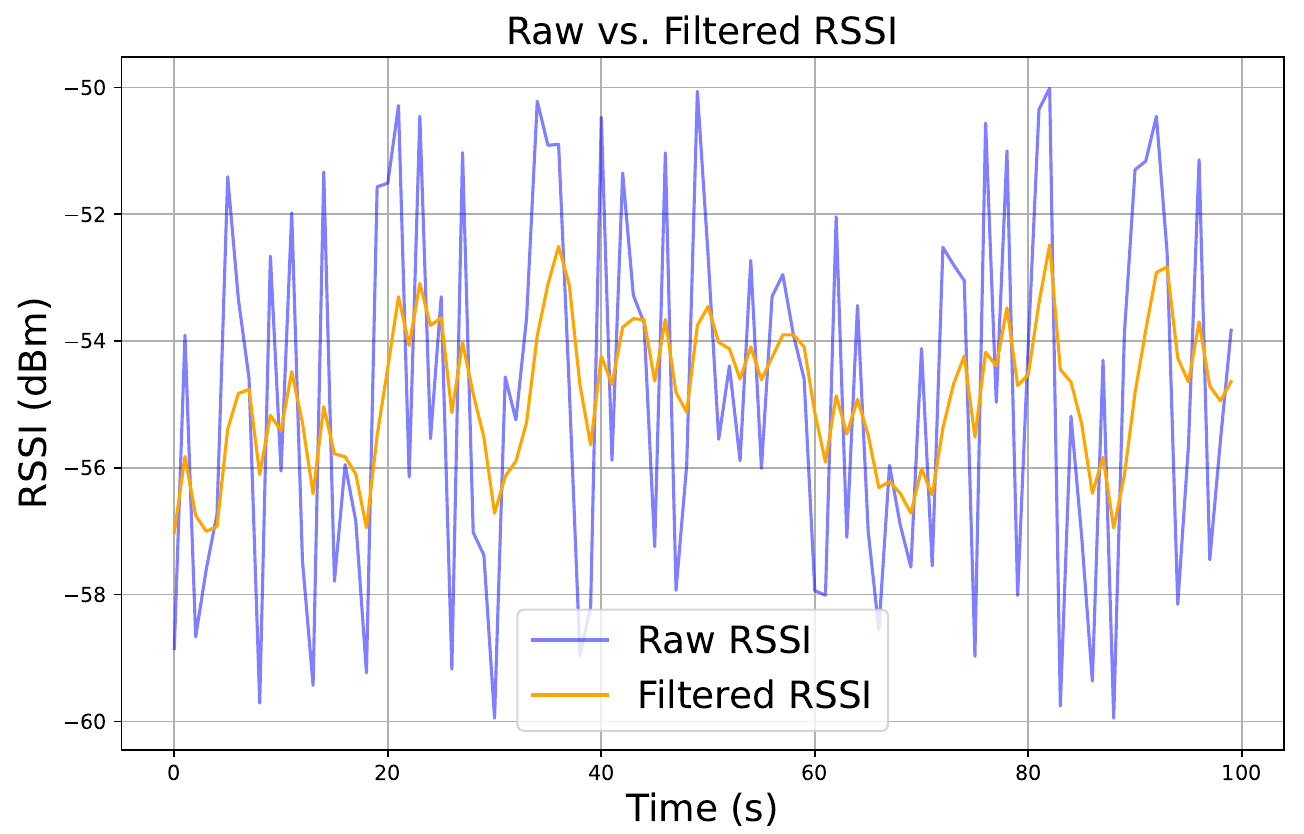}
\DeclareGraphicsExtensions.
\caption{ RSSI values before and after applying the Kalman filter.}
\label{fig_sim}
\vspace{-3mm}
\end{figure}

\begin{figure}[!t]
\centering
\includegraphics[width=.4\textwidth]{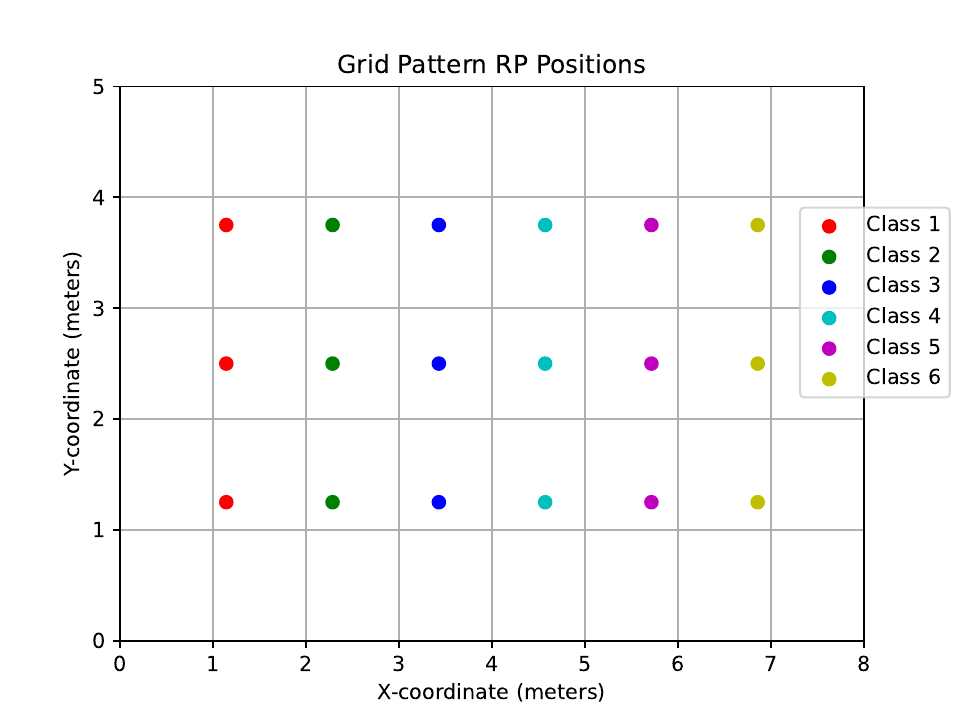}
\DeclareGraphicsExtensions.
\caption{The spatial distribution of RPs in a grid pattern, with different classes represented by distinct colors.}
\label{fig_sim}
\vspace{-3mm}
\end{figure}

\begin{figure}[t]
\centering
\includegraphics[width=.45\textwidth]{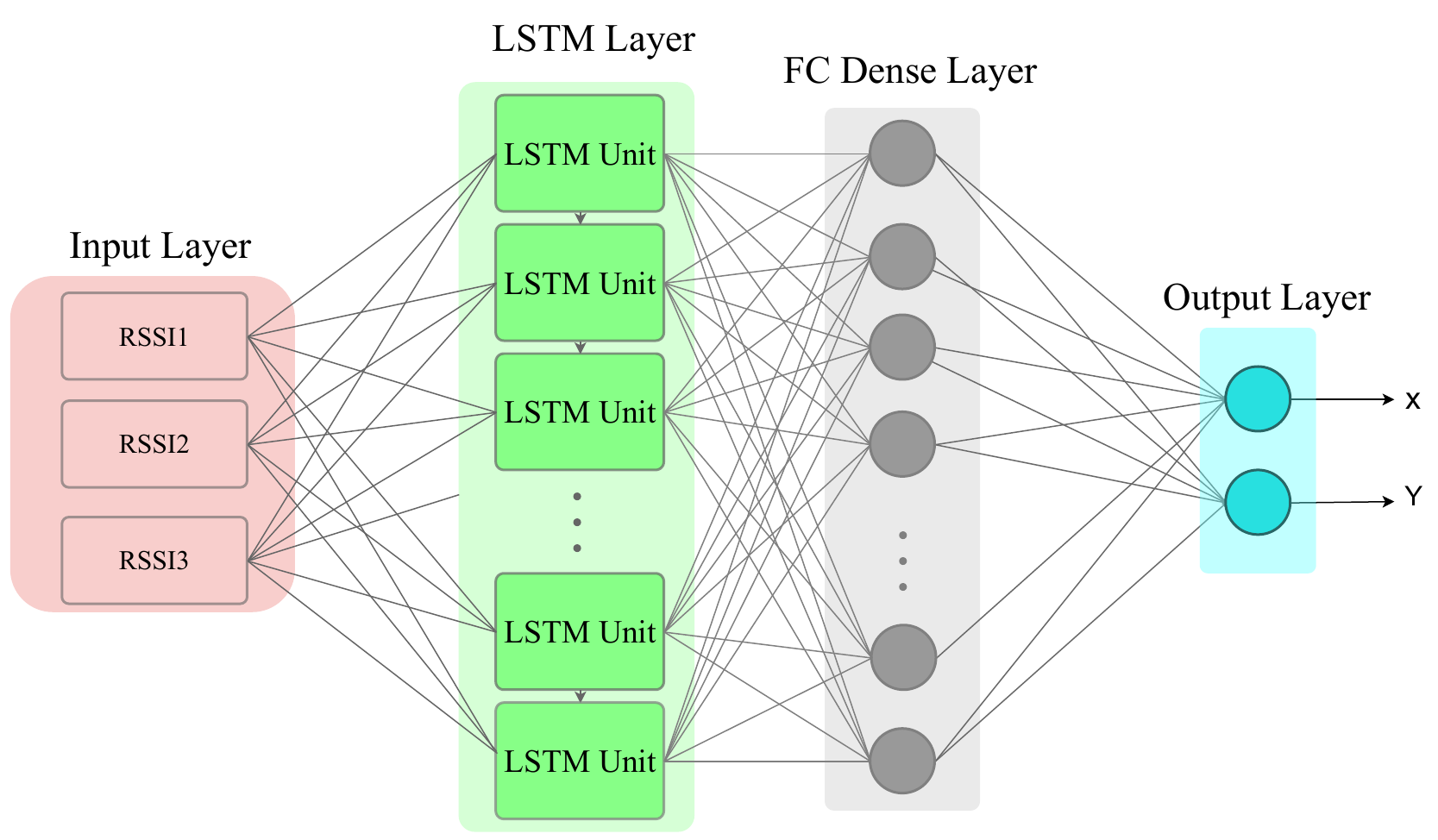}
\DeclareGraphicsExtensions.
\caption{DNN model with an LSTM layer, followed by two dense layers.}
\label{fig_sim}
\vspace{-5mm}
\end{figure}
\begin{table}[t]
\centering
\caption{ANN Model Architecture and Training Configuration}
\begin{tabular}{|c|c|}
\hline
\textbf{Component} & \textbf{Value} \\
\hline
LSTM Units & 32 \\
\hline
Input Shape & (n\_signal\_source, n\_features) \\
\hline
Dense Units & 16 \\
\hline
Activation Function & ReLU \\
\hline
Dense Units & 2 \\
\hline
Optimizer & Adam \\
\hline
Loss Function & Mean Squared Error (MSE) \\
\hline
Epochs & 200 \\
\hline
Batch Size & 32 \\
\hline
\end{tabular}
\end{table}
  1. \textbf{Input Layer:}
        \begin{itemize}
            \item Dimensionality: The input layer receives sequences of RSSI values, each containing measurement from the AP and BLE beacons.
            \item Shape: The input shape is defined by the number of signal sources and the number of features per signal source (i.e., RSSI value in this case).
        \end{itemize}
     2. \textbf{LSTM Layer:}
        \begin{itemize}
            \item Description: The LSTM layer is responsible for capturing temporal patterns and dependencies present in the sequence of RSSI measurements. LSTM is chosen for its ability to mitigate the vanishing gradient problem and effectively model long-range dependencies.
            \item Units: The LSTM layer contains a configurable number of LSTM units, which determine the network's capacity to learn complex temporal relationships.
        \end{itemize}
     3. \textbf{Fully Connected (FC) Dense Layers:}
        \begin{itemize}
            \item Description: Following the LSTM layer, the ANN employs two FC dense layers to transform the learned representations from the LSTM layer and produce the final output.
            \item First Dense Layer:
                \begin{itemize}
                    \item Units: The first dense layer consists of units designed to apply a Rectified Linear Unit (ReLU) activation function. This introduces non-linearity and helps in learning intricate relationships.
                \end{itemize}
            \item Second Dense Layer (Output Layer):
                \begin{itemize}
                    \item Units: The output layer has two units, corresponding to the position to be estimated.
                \end{itemize}
        \end{itemize}
     4. \textbf{Activation Functions:}
        \begin{itemize}
            \item LSTM Layer: No explicit activation function is required for the LSTM layer, as it is capable of learning complex temporal patterns inherently.
            \item First Dense Layer: The ReLU activation function is used to introduce non-linearity and enhance the model's representational power.
        \end{itemize}
     5. \textbf{Model Compilation:}
        \begin{itemize}
            \item Optimizer: The model is compiled using the Adam optimizer, renowned for its adaptive learning rate and effective gradient-based optimization. Its adaptiveness contributes to efficient convergence by dynamically tuning the learning rate based on the historical parameter gradients. Adam integrates the advantages of both momentum and RMSProp optimization techniques. Momentum aids in accelerating convergence by incorporating a running average of prior gradients, whereas RMSProp fine-tunes learning rates for each parameter by adjusting them in inverse proportion to the square root of the mean of past squared gradients.
            \item Loss Function: The chosen loss function is the mean squared error (MSE), which quantifies the difference between predicted and true coordinates.
        \end{itemize}
    
     6. \textbf{Training Procedure:}
        \begin{itemize}
            \item Training Data: The model is trained using a subset of the provided dataset. The training data includes sequences of RSSI measurements as inputs and corresponding ground truth positions as targets.
            \item Epochs and Batch Size: The training process involves iterating through the training data for a predetermined number of epochs. The dataset is divided into batches of a specified size, enabling efficient gradient updates.
        \end{itemize}
    
     7. \textbf{Prediction and Evaluation:}
        \begin{itemize}
            \item Prediction: After training, the trained model is utilized to predict the position for sequences of RSSI measurements from unseen instances.
            \item Evaluation: The Euclidean distance error is calculated between the predicted and actual position for the test instances, providing a quantitative measure of the model's performance.
        \end{itemize}
\begin{algorithm}[t]
\caption{RNN Algorithm for Position Estimation}
\label{alg:rnn}
\textbf{Input:} Training data \texttt{train\_X}, \texttt{train\_y}, LSTM units \texttt{n\_units} \\
\textbf{Output:} Position
\begin{algorithmic}[1]
\STATE Initialize RNN model
\STATE Define number of signal source $n_{signal\_source}$, features $n_{features}$
\STATE Reshape training data to match input dimensions
\STATE Create RNN model with LSTM layer and dense layers
\STATE Compile the model with optimizer and loss function
\STATE Train the model using training data
\STATE Initialize array for predicted positions \texttt{pred\_y}
\FOR{each test sample \texttt{test\_sample} in \texttt{test\_X}}
    \STATE Make a prediction using the trained RNN model
    \STATE Store predicted position in \texttt{pred\_y}
\ENDFOR
\STATE Calculate position estimation error using Euclidean distance
\end{algorithmic}
\end{algorithm}
Algorithm 2 presents the above-explained ANN model training and position estimation.
\section{Experimentations and Results}
In this section, we evaluate the performance of our proposed algorithm in comparison to existing works \cite{Access17}, \cite{Access16},  and \cite{Access18}.

We consider a room of size 4mx7m and place three ESP32 development kits, among which two act as the BLE beacons and one plays the role of Wi-Fi AP. We organize 18 RPs in a grid pattern, arranging them in a 6x3 layout with horizontal and vertical spacing to effectively cover the designated area. Utilizing these RPs, we then solve the optimization problem (2), which results in the locations (1.5, 1.5), (4, 2.5), (7, 1.5) for the BLE beacons and Wi-Fi AP.

We estimate the location of the device using our proposed Algorithm 2. Fig. 6 shows the real and estimated position of this device within the room. As can be observed, the RNN model successfully captures the spatial patterns present in the RSSI data, allowing it to accurately predict the positions of objects within the indoor environment. Notably, the model's predictions exhibit a strong correlation with the actual positions, as indicated in this figure. 

We also compare our proposed algorithm with three existing works \cite{Access16}, \cite{Access17} and \cite{Access18}. Table II summarizes the average error (in m) obtained by our proposed algorithm in comparison with the other works.  The results indicate that our novel indoor positioning algorithm surpasses all three aforementioned works, achieving remarkable accuracy with an average error of 61.29 cm. In particular, our proposed algorithm enhances the error by 56\% comparing to \cite{Access17}. This achievement is obtained by the implementation of a Kalman filter and the utilization of the KNN and RNN architectures, all of which collectively contribute to the exceptional performance of our proposed algorithm.
\begin{figure}[t]
\centering
\includegraphics[width=.4\textwidth]{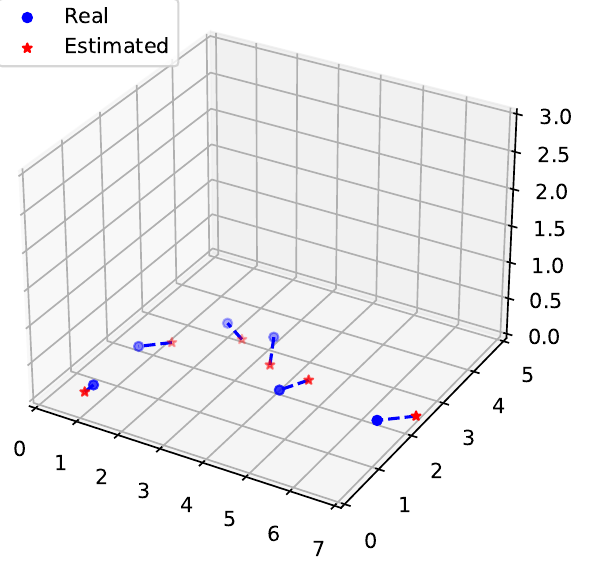} \vspace{-5mm}
\DeclareGraphicsExtensions.
\caption{Actual and estimated locations within an indoor environment.}
\vspace{-2mm}
\label{fig_sim}
\end{figure}
\begin{table}[t]
\centering
\caption{Comparison of Indoor Positioning Accuracy}
\begin{tabular}{lcccc}
\toprule
Paper & Technology & Average Error (meters)\\
\midrule
Current  & Wi-Fi + BLE & 0.6129 \\
\cite{Access16} & BLE & 3\\
\cite{Access17} & Wi-Fi & 1.39\\
\cite{Access18} & RF & 2.8\\
\bottomrule
\end{tabular}
\end{table}

\section{Conclusion}
In this paper, we proposed a highly accurate indoor positioning algorithm leveraging the capabilities of the ESP32 system on chip. We harnessed the power of both Wi-Fi and BLE to construct a comprehensive dataset of RSSI measurements. We first formulated an optimization problem to determine the location of a WiFi AP and two BLE beacons when we aimed to maximize the overall dissimilarity in RSSI values among the RPs. We then collected RSSI values and recognized the intrinsic challenge of dealing with noisy RSSI measurements. In response, we incorporated a Kalman filter to smooth out the fluctuation and  classified the acquired data into distinct classes using the KNN algorithm. We further developed an RNN model, enriched with an LSTM layer, for accurate and robust estimations of device’s positions. Through testbed experiments using ESP32, we finally evaluated the accuracy of our proposed algorithm. Our proposed algorithm can accurately  estimate the position with an average error of 61.29 cm, which surpasses existing works by 56\%. This exceptional performance underscored the potential of our proposed algorithm in the realm of indoor positioning.
\ifCLASSOPTIONcaptionsoff
  \newpage
\fi



%
\bibliographystyle{IEEEtran}
\bibliography{ref.bib}

%





\end{document}